\documentclass[preprint,showpacs,preprintnumbers,amsmath,amssymb,double-spaced]{revtex4}
\usepackage{graphicx}
\usepackage{dcolumn}
\usepackage{bm}

\begin{document}
\title{Entanglement in a Dimerized Antiferromagnetic Heisenberg Chain }

\author{Xiang Hao}

\author{Shiqun Zhu\footnote{Corresponding author, E-mail: szhu@suda.edu.cn}}

\affiliation{School of Physical Science and Technology, Suzhou
University, Suzhou, Jiangsu 215006, People's Republic of China}

\begin{abstract}

The entanglement properties in an antiferromagnetic dimerized
Heisenberg spin-$1/2$ chain are investigated. The entanglement gap,
which is the difference between the ground-state energy and the
minimal energy that any separable state can attain, is calculated to
detect the entanglement. It is found that the entanglement gap can
be increased by varying the alternation parameter. Through thermal
energy, the witness of the entanglement can determine a
characteristic temperature below that an entangled state can be
obtained. The entanglement detected by the energy can provide a
lower bound for that determined by the concurrence. If the
alternation parameter is smaller than a critical value, there is
always no inter-dimer entanglement in the chain.

\end{abstract}


\maketitle
\newpage
\section{Introduction}

The quantum entanglement is considered as key resources of quantum
information processing [1, 2]. The property of entanglement plays
an essential role in understanding and quantifying the physical
systems [3-9]. In recent years, some useful measures of
entanglement have been proposed. The entropy of entanglement is
used to qualify the entanglement of pure states [3] while the
entanglement of formation is one measure for mixed states [4]. The
entanglement in optical systems has been theoretically analyzed
and observed [5-9]. For further developing the experimental
detection, the separability criterion [10, 11] was suggested. On
this basis, the thermal energy [12, 13], the magnetization [14]
and the susceptibility [15] have been used as entanglement witness
[16] to detect the entanglement. If the expected value of the
witness is negative, an entangled state could be obtained. These
measurements can provide an intuitive way to evaluate the quantum
entanglement. With the help of these measurements, it is possible
to find some entangled systems suitable for quantum computation
and quantum communication [17]. Recently, much interest has been
focused on the entanglement in solid-state systems [18-27]. The
spin chain is one kind of entangled systems in condensed matters.
The entanglement has been observed experimentally in Heisenberg
spin chains [18]. The dependence of entanglement on the external
magnetic field [19] and anisotropy [20] has been investigated. The
frustration [21, 22] and arbitrary spin-$s$ Heisenberg chain [23,
24] were considered. Some interesting effects were discussed
concerning the relation of entanglement with correlation [25] and
quantum phase transition [26, 27]. The dimerized spin chain is
another essential sort of spin models in many real solids. It is
often used as a model to explain thermodynamical properties of
many substances [28]. It is interesting to note that there is the
same structure of phase diagrams in the antiferromagnetic
dimerized Heisenberg spin-$1/2$ and uniform spin-$1$ chains [29].
The model belongs to quasi one-dimensional magnets of spin-ladder
systems with even number of coupled spin chains [30]. Therefore,
it is necessary to evaluate and detect the quantum entanglement in
alternating spin chains.

In this paper, the entanglement in an antiferromagnetic dimerized
Heisenberg spin chain is investigated. In Sec. II, the energy is
introduced as a witness for the detection of the dimer
entanglement. The characteristic temperature for the presence of
the dimer entanglement is derived. The entanglement gap is
introduced to evaluate the entanglement for a chain with large
number of spin dimers. The upper bound to the entanglement gap is
derived. In Sec. III, the relation of the entanglement witness and
concurrence is deduced. As an example, the four-spin dimerized
chain with two dimers is analyzed in detail. The limit cases are
discussed when the alternation parameter equals to zero or one. A
brief discussion concludes the paper.

\section{Witness For Dimer Entanglement}

The Hamiltonian of the dimerized Heisenberg spin chain can be
written as [28, 29]
\begin{equation}
\label{eq:1} H=J\sum_{i=1}^{L}[1-(-1)^{i}\delta]\vec{S}_{i} \cdot
\vec{S}_{i+1}
\end{equation}
where $\vec{S}_{i}=\frac 12\vec{\sigma}_{i}$ is the $i$th spin
vector. The number of spins $L$ is even and the periodic boundary
condition $L+1=1$ is assumed. At temperature $T$, the thermal
equilibrium state is $\rho=\exp(-H\beta)/Z$ where
$Z=\mathrm{Tr}[\exp(-H\beta)]$ is the partition function and
$\beta=1/kT$. For the convenience, the Boltzmann constant $k$ is
assumed to be one. The values of the exchange coefficient $J>0$ and
$J<0$ correspond to the antiferromagnetic and ferromagnetic cases
respectively. The parameter $\delta$ denotes the alternating ratio
of exchange interactions. There is an equivalent expression of the
Hamiltonian [28, 29] $H=J^{'}\sum(\vec{S}_{2i-1}\cdot
\vec{S}_{2i}+\alpha\vec{S}_{2i}\cdot \vec{S}_{2i+1})$ with
$J=J^{'}\frac {1+\alpha}2$ and $\delta=\frac {1-\alpha}{1+\alpha}$
where $\alpha$ is the alternation parameter. Such dimerized
Heisenberg spin chain is schematically illustrated in Fig. 1. The
spin chain is constructed by the dimers of the number $d=L/2$. Thus,
the Hamiltonian can be described as $H=\sum_{d=1}^{L/2}H_{d}$. In
Fig. 1, the spin dimer is labelled by an elliptical box. The solid
line represents the dimer interaction $J^{'}$ and the dash line
denotes the inter-dimer interaction $\alpha J^{'}$. In the following
discussions, the antiferromagntic case of $J^{'}=1$ and $0\leq
\alpha \leq1$ is considered. It is easily found that the ground
state at $\alpha=0$ is a tensor product of a singlet state. When
$\alpha \neq 0$, the ground state cannot be expressed as a tensor
product of each dimer state. For the simplest case of $d=2$, the
ground state at $\alpha=1$ can be written as $|\psi_0\rangle=\frac
1{\sqrt{12}}[(|1100\rangle+|0011\rangle+|1001\rangle+|0110\rangle)-2(|1010\rangle+|0101\rangle)]$.
Here $|1\rangle,|0\rangle$ are assumed to be the eigenstates of the
pauli operator $\sigma^{z}$ with the eigenvalues $\pm 1$. Therefore,
it is clear that there exists the transition of the ground state
$|\psi_0\rangle$ if $\alpha$ is varied from zero.

On the basis of the separability criterion, there exists a minimal
separable energy $E_{sep}$ [12, 13]. That is, $E_{sep}$ is the
minimal energy that any separable state can arrive at. For the
antiferromagnetic dimerized Heisenberg chain, the pure separable
state for the minimal energy can be analyzed by the standard
symmetry methods [24, 31] and expressed as
$|\psi_{sep}\rangle=\frac
1{2^L}\prod_{i=1}^{L}(|0\rangle_i-|1\rangle_i)\otimes(|0\rangle_{i+1}+|1\rangle_{i+1}).$
The minimal separable energy is $E_{sep}=\langle
\psi_{sep}|H|\psi_{sep}\rangle=-\frac {L(1+\alpha)}8$. For the
detection of the dimer entanglement, the entanglement gap $g_E$,
which is the difference between the minimal separable energy
$E_{sep}$ and the ground-state energy $E_{0}$, can be introduced.
The upper bound of the entanglement gap can be derived. The lower
bound to the ground state energy of the Hamiltonian
\begin{equation}
\label{eq:2} H=J^{'}\sum(\vec{S}_{2i-1}\cdot
\vec{S}_{2i}+\alpha\vec{S}_{2i}\cdot \vec{S}_{2i+1})
\end{equation}
can be estimated. By the variation principle, one can show that
the ground state energy of the Hamiltonian $H$ can be written as
\begin{equation}
\label{eq:3} E_0(H)\geq E_0(H_1) + \alpha E_0(H_2).
\end{equation}
Where $E_0(H_1)$ is the ground state energy of the Hamiltonian
$H_1=\sum(\vec{S}_{2i-1}\vec{S}_{2i})=\vec{S}_1\vec{S}_2+\vec{S}_3\vec{S}_4+......$,
while $E_0(H_2)$ is the ground state energy of the Hamiltonian
$H_2=\sum(\vec{S}_{2i}\vec{S}_{2i+1})=\vec{S}_2\vec{S}_3+\vec{S}_4\vec{S}_5+......$.
Since
\begin{equation}
\label{eq:4}\vec{S}\vec{S}=\frac {1}{4}\left(\begin{array}{cccc}
1&0&0&0\\
0&-1&2&0\\
0&2&-1&0\\
0&0&0&1
\end{array}\right),
\end{equation}
one has $E_0(H_1)=E_0(H_2)=-\frac{3}{8} L$. Thus for the ground
state $E_0(H)\geq-\frac{3}{8}L(\alpha+1)$. The expression of the
entanglement gap $g_E$ and its upper bound can be given by
\begin{equation}
\label{eq:5} g_E=\frac {E_{sep}-E_0}{L}\leq \frac{1}{4}(\alpha+1).
\end{equation}
If the entanglement gap is nonzero, the entanglement could be
detected experimentally below a certain temperature. To some
degree, the entanglement gap can be regarded as a useful indicator
of the existence of entanglement. The entanglement at a finite
temperature can be detected more easily if the entanglement gap is
large [12]. For the dimerized chain with a large number of dimers,
the ground-state energy $E_0$ can be numerically calculated. The
dependence of the entanglement gap on the alternation parameter is
plotted in Fig. 2. It is shown that the entanglement gap is
decreased with the increase of the alternation $\alpha$ and
reaches a minimal value at about $\alpha=0.7$, and then increased
again until $\alpha=1.0$. Here the dimer interaction is chosen as
$J'=1$. It is found that the entanglement gap $g_E$ can be
increased when $\alpha \neq 0.7$. Meanwhile, the entanglement
witness at finite temperature $T$ can be introduced as [12, 13]
\begin{equation}
\label{eq:6} W=E-E_{sep}
\end{equation}
where $E=\sum_{d}\mathrm{Tr}(\rho H_{d})$ is the energy at the
thermal state $\rho$. If the value $W<0$, an entangled state is
obtained. The value of $W>0$ means that it is a separable
(unentangled) state. The entanglement witness $W$ as a function of
the temperature $T$ and the alternating parameter $\alpha$ is
plotted in Fig. 3. Figure 3(a) is a three-dimensional plot of the
entanglement witness $W$ as a function of $\alpha$ and $T$. From
Fig. 3(a), it is seen that the value of $W$ is increased to a
maximum and then decreased when $\alpha$ is increased. The value
of $W$ is increased from negative to the positive when the
temperature $T$ is increased. That is, the thermal states are
changed from the entangled states to the separable states as $T$
is increased. The contour of $W$ as a function of the
characteristic temperature $T_{c}$ and the alternation parameter
$\alpha$ is illustrated in Fig. 3(b). The value of $T_{c}$ is
located on the contour map of $W=0$. The contour map of $W<0$
corresponds to the entangled thermal states. It is found that both
entanglement-detecting methods are almost equivalent to each
other.

\section{Relation of Entanglement Witness and Concurrence}

Although some of the real solids is composed of a large number of
spins, many properties, like thermal and magnetic properties can be
efficiently studied by the model of small number of spins. It is
necessary to analyze the entanglement properties of the dimerized
chain with small number of dimers. In the Hilbert space
$\{|11\rangle,|10\rangle,|01\rangle,|00\rangle \}$, the reduced
density matrix of one dimer $\rho_d$ can be written as
\begin{equation}
\label{eq:7} \rho_d=\left(\begin{array}{cccc}
u_{d}&0&0&0\\
0&w_{d}&t_{d}&0\\
0&t_{d}&w_{d}&0\\
0&0&0&u_{d}
\end{array}\right)
\end{equation}
where the elements are expressed as $u_{d}=\frac 14+\frac
1{12}\mathrm{Tr}[\rho_{d}\vec{\sigma}\cdot \vec{\sigma}]$,
$w_{d}=\frac 14-\frac 1{12}\mathrm{Tr}[\rho_{d}\vec{\sigma}\cdot
\vec{\sigma}]$ and $t_{d}=\frac
1{6}\mathrm{Tr}[\rho_{d}\vec{\sigma}\cdot \vec{\sigma}]$. The
concurrence of the system is given by
$C=\max\{0,2\lambda_1-\sum_i\lambda_i\}$ where $\lambda_i$ are the
square roots of eigenvalues of the matrix $R=\rho(\sigma^y\otimes
\sigma^y)\rho(\sigma^y\otimes \sigma^y)$ in decreasing order and
$\lambda_1$ is the maximum one [4]. From the definition of the
concurrence $C$, the dimer entanglement $C$ of the alternating
Heisenberg chain can be calculated as
$C=2\max\{0,|u_{d}|-|t_{d}|\}=\frac
12\max\{0,-\mathrm{Tr}[\rho_{d}\vec{\sigma}\cdot
\vec{\sigma}]-1\}$. Here $\mathrm{Tr}(\rho_{d}\vec{\sigma}\cdot
\vec{\sigma})<0$ is the correlation function which is
monotonically increased with the temperature $T$ [23]. According
to Eq. (6), the witness $W$ can also be expressed by the
correlation function
\begin{equation}
\label{eq:8} W=\frac
{L(1+\alpha)}8[1+\mathrm{Tr}(\rho_{d}\vec{\sigma}\cdot
\vec{\sigma})]
\end{equation}
Thus, the relation of the entanglement witness $W$ and concurrence
$C$ can be expressed as
\begin{equation}
\label{eq:9} C\geq \max\{0,-\frac {4W}{(1+\alpha)L}\}.
\end{equation}
Here the equality can be achieved if the alternation parameter
$\alpha=0$ or $1$. The item of $\max\{0,-\frac
{4W}{(1+\alpha)L}\}$ denotes the entanglement witnessed by the
thermal energy. From Eq. (9), the detection of the entanglement
can offer a lower bound for the concurrence $C$.

In the limit case of $\alpha=0$ or $\alpha=1$, the dimerized
Heisenberg spin chain is reduced to a single dimer or an isotropic
Heisenberg chain. For the case of $\alpha=1$, the concurrence for
any one dimer is just the entanglement between two nearest
neighboring spins. It can be written by $C_{i,i+1}(\alpha=1)=\frac
12\max\{0,-K_{i,i+1}-1\}$, where
$K_{i,i+1}=\mathrm{Tr}[\rho\vec{\sigma}_{i}\cdot
\vec{\sigma}_{i+1}]$ is the correlation function for any two
nearest neighboring spins. According to Eq. (6), the entanglement
witness can be expressed by $W(\alpha=1)=E+\frac L4$. Here the
minimal separable energy is $E_{sep}=-\frac L4$. For such an
isotropic spin chain, the thermal energy is
$E(\alpha=1)=\sum_{i}^{L}\mathrm{Tr}(\rho \vec{S}_{i}\cdot
\vec{S}_{i+1})=\frac L4 K_{i,i+1}$. It is easily seen that the
concurrence $C_{i,i+1}(\alpha=1)=\max\{0, \frac {-2W}{L}\}$ [24].
When $\alpha=0$, it is easily seen that there is no inter-dimer
interaction, i. e., the correlation function
$K_{2i,2i+1}(\alpha=0)=0$. The correlation function $K_{2i-1,2i}$
can still exist in the chain. The entanglement witness for
$\alpha=0$ can also be written by
$W(\alpha=0)=\sum_{i}^{L/2}\mathrm{Tr}(\rho \vec{S}_{2i-1}\cdot
\vec{S}_{2i})+\frac L8=\frac L8 (K_{2i-1,2i}+1)$. Similarly, the
concurrence for a single dimer is $C_{2i-1,2i}(\alpha=0)=\max\{0,
-\frac 12(K_{2i-1,2i}+1)\}=\max \{0, \frac {-4W}{L} \}$.
Therefore, the equality in Eq. (9) is obtained when the
alternation parameter is in the limit of $\alpha=0$ or $\alpha=1$.

As a simple example, the antiferromagnetic dimerized Heisenberg
chain with two dimers is analyzed. Because the Hamiltonian satisfies
$[H, \sum_iS_{i}^{z}]=0$, the total spin number $S$ is conserved.
The eigenvalues $E_i$ of the Hamiltonian can be written as
\begin{eqnarray}
\label{eq:10}
E_1=E_2=E_3=E_{10}=E_{13}=\epsilon_1=(1+\alpha)/2, \\
E_4=E_9=E_{14}=\epsilon_2=-(1+\alpha)/2,\nonumber \\
E_5=E_7=E_{15}=\epsilon_3=(1-\alpha)/2,\nonumber \\
E_6=E_8=E_{16}=\epsilon_4=-(1-\alpha)/2,\nonumber \\
E_{11}=\epsilon_5=-(1+\alpha)/2+\sqrt{\alpha^2-\alpha+1},\nonumber \\
E_{12}=\epsilon_6=-(1+\alpha)/2-\sqrt{\alpha^2-\alpha+1},\nonumber
\end{eqnarray}
The corresponding eigenstates $|\psi_{i}\rangle$ can also be expressed by
\begin{eqnarray}
\label{eq:11}
|\psi_1\rangle&=&|1111\rangle; |\psi_2\rangle=|0000\rangle; \\
|\psi_3\rangle&=&\frac
12(|0111\rangle+|1101\rangle+|1011\rangle+|1110\rangle);
|\psi_4\rangle=\frac
12(|0111\rangle+|1101\rangle-|1011\rangle-|1110\rangle);\nonumber\\
|\psi_5\rangle&=&\frac
12(|0111\rangle-|1101\rangle+|1011\rangle-|1110\rangle);
|\psi_6\rangle=\frac
12(|0111\rangle-|1101\rangle-|1011\rangle+|1110\rangle);\nonumber\\
|\psi_7\rangle&=&\frac
1{\sqrt2}(|0011\rangle-|1100\rangle);|\psi_8\rangle=\frac
1{\sqrt2}(|1001\rangle-|0110\rangle);|\psi_9\rangle=\frac
1{\sqrt2}(|1010\rangle-|0101\rangle);\nonumber\\
|\psi_{10}\rangle&=&\frac
1{\sqrt6}(|1010\rangle+|0101\rangle)+|1001\rangle+|0110\rangle+|1100\rangle+|0011\rangle);\nonumber\\
|\psi_{11}\rangle&=&\frac 1{\sqrt
{2(1+x_{+}^2+y_{+}^2)}}[x_{+}(|1100\rangle+|0011\rangle)+y_{+}(|1010\rangle+|0101\rangle)+|1001\rangle+|0110\rangle];\nonumber
\\
|\psi_{12}\rangle&=&\frac 1{\sqrt
{2(1+x_{-}^2+y_{-}^2)}}[x_{-}(|1100\rangle+|0011\rangle)+y_{-}(|1010\rangle+|0101\rangle)+|1001\rangle+|0110\rangle];\nonumber
\\
|\psi_{13}\rangle&=&\frac
12(|1000\rangle+|0010\rangle+|0100\rangle+|0001\rangle);|\psi_{14}\rangle=\frac
12(|1000\rangle+|0010\rangle-|0100\rangle-|0001\rangle);\nonumber \\
|\psi_{15}\rangle&=&\frac
12(|1000\rangle-|0010\rangle+|0100\rangle-|0001\rangle);|\psi_{16}\rangle=\frac
12(|1000\rangle-|0010\rangle-|0100\rangle+|0001\rangle)\nonumber
\end{eqnarray}
where the parameters of $x_{\pm}$ and $y_{\pm}$ are given by
$x_{\pm}=-[(1-\alpha)\pm \sqrt{\alpha^2-\alpha+1}]$ and
$y_{\pm}=-\alpha \pm \sqrt{\alpha^2-\alpha+1}$. At finite
temperature $T$, the correlation function
$K=\mathrm{Tr}(\rho_{d}\vec{\sigma}\cdot \vec{\sigma})$ can be
calculated as
\begin{equation}
\label{eq:12} K=-3+\frac
1Z(20e^{-\epsilon_1\beta}+6e^{-\epsilon_2\beta}+12e^{-\epsilon_3\beta}+6e^{-\epsilon_4\beta}+\frac
{6x_{-}^2}{1+x_{-}^2+y_{-}^2}e^{-\epsilon_5\beta}+\frac
{6x_{+}^2}{1+x_{+}^2+y_{+}^2}e^{-\epsilon_6\beta})
\end{equation}
From Eqs. (8)-(12), the concurrence $C$ and the entanglement
witness $W$ of one dimer can be easily calculated.

It is also known that there is inter-dimer interaction in the
antiferromagnetic dimerized Heisenberg spin-$1/2$ chain. The
concurrence $C_{2i,2i+1}$ between two spins $2i$ and $2i+1$
denotes the inter-dimer entanglement. It is induced from the the
inter-dimer exchange coupling $\alpha J'$. The reduced density
matrix between any two spins has the same structure as that given
by Eq. (7). Therefore, the entanglement $C_{i,j}$ between any two
spins $i$ and $j$ is also dependent on the correlation function
$K_{i,j}$, which can be expressed by
\begin{equation}
\label{eq:13} C_{i,j}=\frac 12\max\{0,-K_{i,j}-1\}.
\end{equation}
The inter-dimer correlation function is
$K_{2i,2i+1}=\mathrm{Tr}[\rho_{2i, 2i+1}\vec{\sigma}\cdot
\vec{\sigma}]$, where $\rho_{2i,2i+1}$ is the reduced density
matrix for inter-dimer two spins. It is clearly seen that there is
the inter-dimer entanglement if the correlation function
$K_{2i,2i+1}<-1$. For such spin chain, the correlation function
between any two spins is increased with the increase of the
temperature [23]. Thus, the inter-dimer entanglement of the
concurrence $C_{2i,2i+1}$ is always decreased from that of the
ground state. When the correlation for the ground state
$K^{0}_{2i,2i+1}\geq -1$, there is no inter-dimer entanglement at
any temperature $T$. As an example, for the case of $L=4$, the
ground state is given by the state $|\psi_{12}\rangle$. The
inter-dimer correlation function for the ground state is obtained
by $K^{0}_{2,3}=-3(1-\frac 2{1+x_{-}^2+y_{-}^2})$. In the limit of
$\alpha=0$, $K^{0}_{2,3} =0$, there is no inter-dimer
entanglement. When the value of $\alpha$ is increased from zero,
the value of inter-dimer correlation function is also decreased
from zero. If $\alpha> 0.5$, $K^{0}_{2,3} < -1$, the inter-dimer
entanglement exists. In the limit of $\alpha=1$, $K^{0}_{2,3} =
-2$, the inter-dimer entanglement reaches the maximum value. The
inter-dimer entanglement of the concurrence $C_{2,3}$ can be
calculated by the correlation function of $K^{0}_{2,3}$. By
numerical calculations, it is found that there is no inter-dimer
entanglement when the alternation parameter is below a critical
value of $\alpha_c$. The critical alternation parameter $\alpha_c$
is plotted in Fig. 4(a) as a function of spin number $L$. It is
seen that the value of $\alpha_c$ increases and then saturates to
a constant of $\alpha^s_c\sim 0.78$ at $L=12$. To show the
saturation effect, the characteristic temperature $T_c$ of $W=0$
is numerically calculated as a function of the alternation
parameter $\alpha$ and is plotted in Fig. 4(b) when the number of
spins $L$ is varied. It is seen that the value $T_{c}$ of $L=10$
(the bullet) keeps almost the same value as that of $L=12$ (the
triangle) for relatively large value of the alternation parameter
$\alpha$.

It is very interesting to note that the entanglement between two
spins depends on the separable distance $|j-i|$ in the
antiferromagnetic dimerized Heisenberg spin-$1/2$ chain. This can
be illustrated by the correlation function $K^{0}_{1,j}$ between
two spins $1$ and $j$ for the ground state. For the chain with
spin number $L=12$, the correlation function
$K^{0}_{1,j}=\mathrm{Tr}[\rho^{0}_{1,j}\vec{\sigma}\cdot
\vec{\sigma}]$ is numerically calculated and plotted in Fig. 5 as
a function of separable distance $|j-1|$. Here $\rho^{0}_{1,j}$ is
the reduced density matrix between spins $1$ and $j$ for the
ground state. In Fig. 5, the dashed horizontal line denotes the
value of $K^{0}_{1,j} = -1$. Below the horizontal line, the
entanglement exists. From Fig. 5, it is clear that the values of
the correlation function is $K^{0}_{1,j}>-1$ when the separable
distance is $|j-1|>2$. According to the relation of the
correlation function and the concurrence, the concurrence
$C_{1,j}$ is always zero for $|j-1|>2$. In the Heisenberg chain,
only $K^{0}_{1,2}$ is smaller than $-1$ while the others
$K^{0}_{1,j(j>2)}>-1$. The entanglement always declines from the
ground state at $T=0$. Therefore, at any temperature $T$, there is
no entanglement between two spins $i$ and $j$ when the separable
distance is $|j-i|>2$ in the antiferromagnetic dimerized
Heisenberg spin-$1/2$ chain.

\section{Discussion}

The entanglement in the antiferromagnetic dimerized Heisenberg
spin-$1/2$ chain is analyzed. The entanglement gap $g_E$ can be
increased if the alternation parameter $\alpha\neq 0.7$. If there is
a large number of dimers in the chain, the characteristic
temperature $T_c$ determined by the entanglement witness is
decreased to a constant value. The entanglement can be detected
below a certain temperature in real solids. The relation of the
witness and the concurrence is also derived. The energy as the
entanglement witness can provide a lower bound for the dimer
entanglement. In the limit of $\alpha=0$ or $1$, both of the witness
and the concurrence are equivalent to detect the entanglement. It is
also found that there is no inter-dimer entanglement when the
alternation parameter is smaller than a critical value. These
results may be helpful for the further study of solid-state quantum
communication and computation.

\vskip 0.4cm {\large \bf Acknowledgements}

The financial support from Specialized Research Fund for the
Doctoral Program of Higher Education (Grant No. 20050285002) is
gratefully acknowledged. It is a pleasure to thank Yinsheng Ling
and Jianxing Fang for their many fruitful discussions about the
topic.

\newpage

{\Large \bf Reference}

1. D. P. DiVincenzo, D. Bacon, J. Kempe, G. Burkard, and K. B.
Whaley, Nature

\ \ \ \ (London) {\bf 408}, 339(2000).

2. N. J. Cerf, M. Bourennane, A. Karlsson, and N. Gisin, Phys. Rev.
Lett. {\bf 88},

\ \ \ \ 127902(2002).

3. C. H. Bennett, H. J. Bernstein, S. Popescu, and B. Schumacher,
Phys. Rev. {\bf A53},

\ \ \ \ 2046(1996).

4. W. K. Wootters, Phys. Rev. Lett. {\bf 80}, 2245(1998).

5. X.-M. Lin, Z.-W. Zhou and G.-C. Guo, Phys. Lett. {\bf A348},
299(2006).

6. W. Jiang, C. Han, P. Xue, L.-M. Duan, and G.-C. Guo, Phys. Rev.
{\bf A69},

\ \ \ \ 043819(2004).

7. Y.-S. Zhang, M.-Y. Ye, and G.-C. Guo, Phys. Rev. {\bf A71},
062331(2005).

8. L. Ye, L.-B. Yu, and G.-C. Guo, Phys. Rev. {\bf A72},
034304(2005)

9. K.-H. Song,Z.-W. Zhou, and G.-C. Guo, Phys. Rev. {\bf A71},
052310(2005)

10. A. Peres, Phys. Rev. Lett. {\bf 77}, 1413(1996).

11. M. Horodecki, P. Horodecki, and R. Horodecki, Phys. Lett. {\bf
A223}, 8(1996).

12. M. R. Dowling, A. C. Doherty, and S. D. Bartlett, Phys. Rev.
{\bf A70}, 062113(2004).

13. G. T\'{o}th, Phys. Rev. {\bf A71}, 010301(R)(2005).

14. X. Wang and P. Zanardi, Phys. Lett. {\bf A301}, 1(2002).

15. M. Wiesniak, V. Vedral, and C. Brukner, New. J. Phys. {\bf 7},
258(2005).

16. M. Lewenstein, B. Kraus, J. I. Cirac, and P. Horodecki, Phys.
Rev. {\bf A62},

\ \ \ \ 052310(2000).

17. Y.-J. Han, Y. Hu, Y.-S. Zhang, and G.-C. Guo , Phys. Rev. {\bf
A72}, 064302(2005).

18. S. Ghosh, T. F. Rosenbanm, G. Aeppli, and S. N. Coppersmith,
Nature

\ \ \ \ (London) {\bf 425}, 48(2003).

19. M. C. Arnesen, S. Bose, and V. Vedral, Phys. Rev. Lett. {\bf
87}, 017901(2001).

20. G. L. Kamta and A. F. Starace, Phys. Rev. Lett. {\bf 88},
107901(2002).

21. C. M. Dawson and M. A. Nielsen, Phys. Rev. {\bf A69},
052316(2004).

22. Z. Sun, X. Wang, and Y. Q. Li, New J. Phys. {\bf 7}, 83(2005).

23. J. Schliemann, Phys. Rev. {\bf A68}, 012309(2003).

24. X. Hao and S. Zhu, Phys. Rev. {\bf A72}, 042306(2005).

25. U. Glaser, H. B\"{u}ttner, and H. Fehske, Phys. Rev. {\bf A68},
032318(2003).

26. T. J. Osborne and M. A. Nielsen, Phys. Rev. {\bf A66},
032110(2002).

27. G. Vidal, J. I. Latorre, E. Rico, and A. Kitaev, Phys. Rev.
Lett. {\bf 90}, 227902(2003).

28. T. Barnes, J. Riera, and D. A. Tennant, Phys. Rev. {\bf B59},
11384(1999).

29. E. Herrling, G. Fischer, S. Matejcek, B. Pilawa, H. Henke, I.
Odenwald, and

\ \ \ \ \ W. Wendl, Phys. Rev. {\bf B67}, 014407(2003).

30. E. Dagotto, Rep. Prog. Phys. {\bf 62}, 1525(1999).

31. T. Eggeling and R. F. Werner, Phys. Rev. {\bf A63},
042111(2001).

\newpage

{\Large \bf Figure Captions}

{\bf Fig. 1.}

The antiferromagnetic alternating Heisenberg spin-$1/2$ chain can
be described by spin dimers. The spin dimer is labelled by an
elliptical box. The solid line represents the dimer interaction
and the dash line denotes the inter-dimer interaction.

{\bf Fig. 2.}

The entanglement gap $g_E$ is plotted as a function the
alternation parameter. The minimal gap is located at about
$\alpha=0.7$.

{\bf Fig. 3.}

The witness $W$ is plotted as a function of the alternation
parameter $\alpha$ and the temperature $T$.

(a). The three dimensional plot of $W$ as a function of $\alpha$
and $T$.

(b). The contour plot of $W$ on the plane of $(T-\alpha)$.

{\bf Fig. 4.}

The characteristic alternation parameter $\alpha_c$ and
temperature $T_c$ are plotted.

(a). $\alpha_c$ is plotted as a function of the number of spins
$L$.

(b). $T_c$ is plotted as a function of $\alpha$ with $L=6(\circ),
10(\bullet), 12(\triangle)$.

{\bf Fig. 5}

The correlation function of the ground state $K^{0}_{1,j}$ is
plotted when the spin separable distance $|j-1|$ is increased from
$1$ to $L-1$. Here spin number $L=12$ and the alternation
parameter $\alpha$ is varied from $\alpha=0.3(\circ)$ to
$\alpha=0.8(\bullet)$ to $\alpha=1(\triangle)$. The dash line
denotes the value $-1$.

\newpage

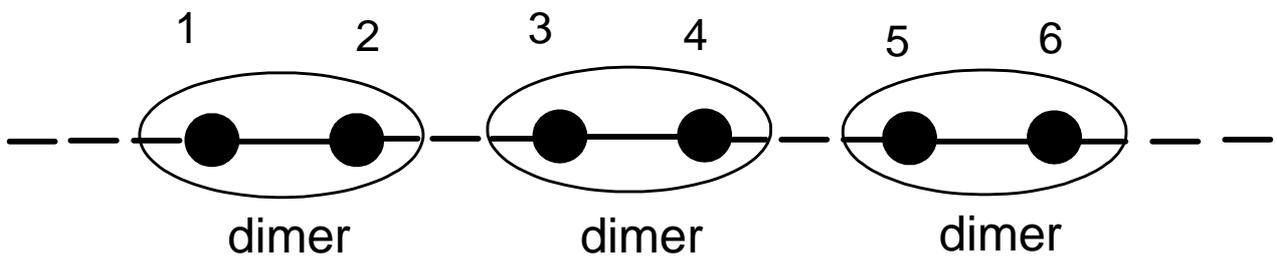
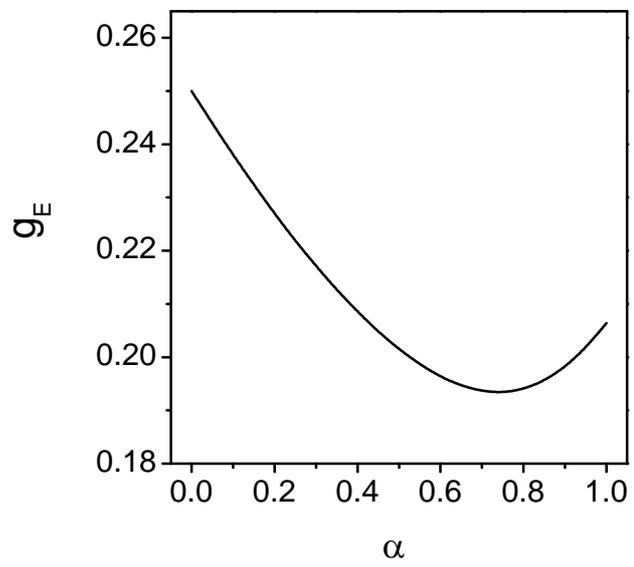
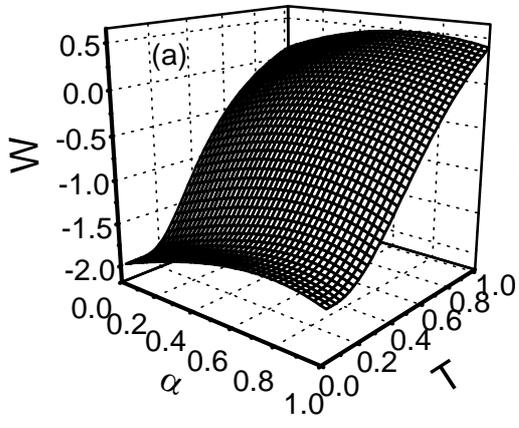
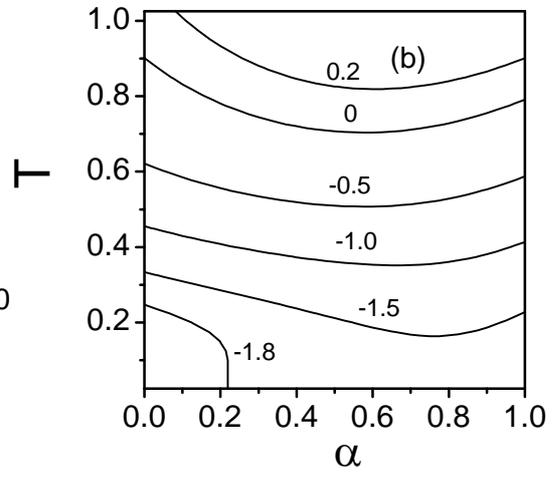
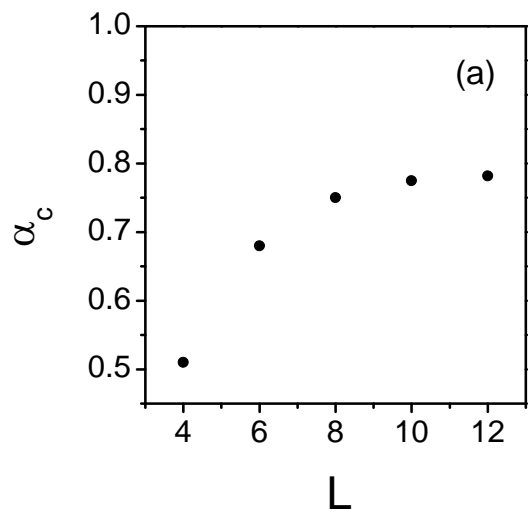
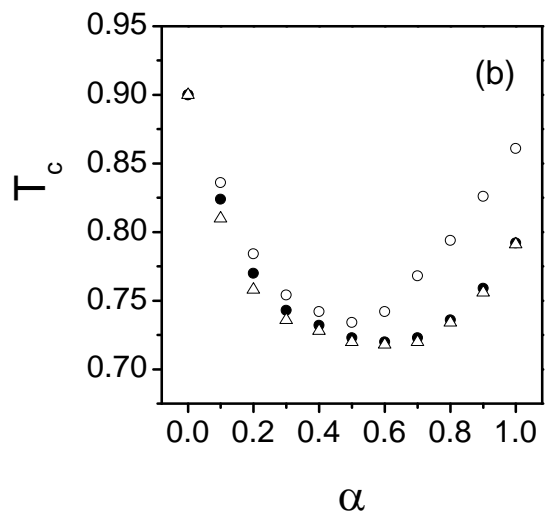
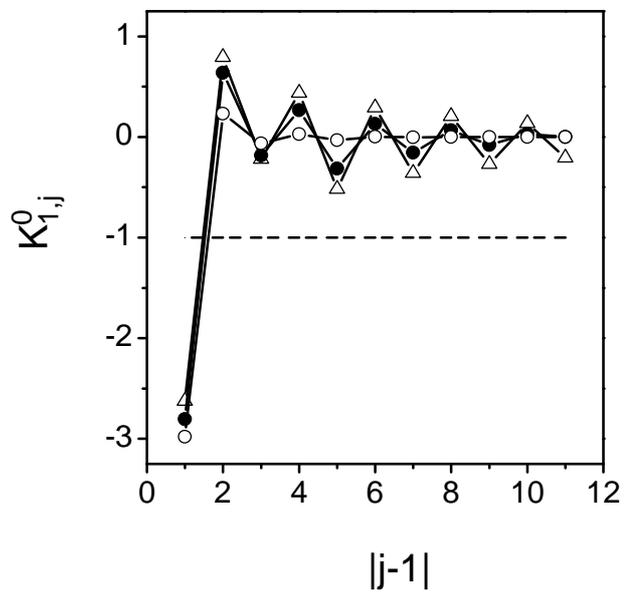
\end{document}